# PCB Coil Design Producing a Uniform Confined Magnetic Field

Peter A. Koss[1], Christopher Crawford[2], Georg Bison[3], Elise Wursten[1], Malgorzata Kasprzak[1], Nathal Severijns[1]

[1]Institute for Nuclear and Radiation Physics, KULeuven, 3001 Heverlee Belgium
[2]Department of Physics, University of Kentucky, Lexington, KY 40506 USA
[3]Paul Scherrer Institute, 5232 Villigen, Switzerland

We present a magnetic field confining coil with a sub $10^{-3}$ field uniformity over a large fraction of the coil. The structure is entirely made out of printed circuit boards (PCB). The PCB design allows to tailor the path of wires to fit the required geometry. We measure the field uniformity with Cesium magnetometers in a field range from $1\mu$T to $10\mu$T. Our application uses such a coil for an atomic magnetometry based current controller.

## I. INTRODUCTION

MANY applications in modern research require a very uniform magnetic field over the experimental volume. Traditionally, a pair of Helmholtz coils or a solenoid would be used. They produce a reasonably uniform field for a coil geometry which has been known and used for many years [1], [2]. Extensions of the Helmholtz coils attempt to improve on the uniformity of the magnetic field by using more than just two coils [3].

Some applications have very stringent requirements on the field uniformity [4]. There one would prefer a $\cos\theta$ or a spherical coil, which typically produces very uniform fields over a large volume of the coil [5]. All of these coils have large fringe fields. They may interact with environmental factors such as nearby high permeability materials, perturbing the actual uniform volume. One may then use either a return yoke or build a field confining coil to guide the stray field and reduce the influence of the environment [6], [7].

We present simulations and measurements of a PCB based field confining coil. The PCB design exploits the precision of modern PCB manufacturing. This allows for tailored wire paths as well as to choose the wire density, i.e. the coil constant. Furthermore, the wire width and thickness can be chosen. Thus, we can adapt the total resistivity of the coil to the used current source. Additionally, the use of PCB headers as connectors makes it easy to open and close this type of field confining coil.

### A. Motivation

Our groups are involved in an international collaboration searching for the neutron electric dipole moment (nEDM) [8], [9]. The experiment uses Ramsey's method of time-separated oscillating fields [10], where neutrons precess in a magnetic field $\vec{H}$ and a parallel, or anti-parallel, electric field $\vec{E}$. These types of measurements require a very stable and uniform magnetic field $\vec{H}_0$. The long-term stability for $\vec{H}_0$ is achieved with a mu-metal shield and active field stabilization [11]. We

Corresponding author: Peter A. Koss (email: peter.koss@kuleuven.be)

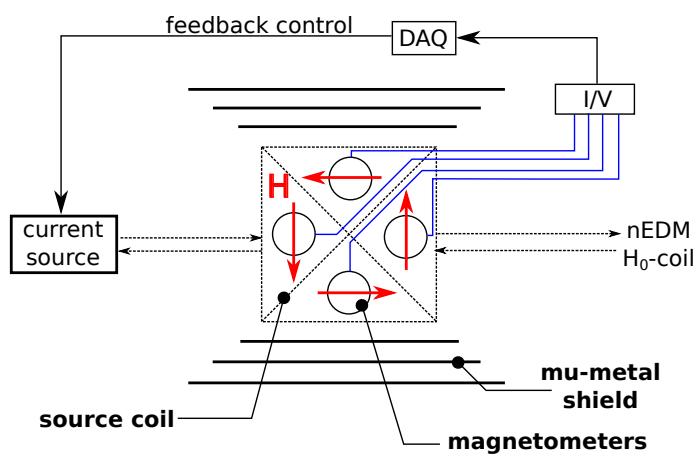

Fig. 1. **Current controller concept.** A low noise current source provides the base current. Changes in this current are registered as changes in the magnetic field in a dedicated coil, the source coil, which is connected in series to the nEDM $H_0$-coil. A magnetometer array in the source coil is able to detect these changes. A dedicated data acquisition system (DAQ) generates the appropriate feedback response for the detected drift.

aim at improving the shielding factor in our experiment by at least a factor 10. Then, the stability will be fundamentally limited by the current source which feeds the $\vec{H}_0$-coil. Thus, the present current source, $10^{-7}$ stability on 17 mA, must be improved. In order to improve on this stability, a current controller based on atomic magnetometry is presently being developed, see Fig.1 [12]. Modern atomic magnetometers easily reach sensitivities on the order of $10^{-8}$ for one second of integration time in the shot noise limit [13], [14]. These sensitivities can be exploited by converting drifts in currents into magnetic field changes [15]. The coil in Fig.1 is able to discriminate external field perturbations from a drift in current. The latter changes the field modulus in all quadrants by the same amount, while an external field will affect each quadrant in a different way.

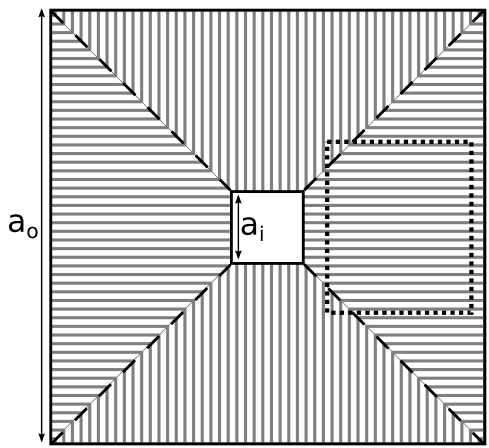

Fig. 2. **Coil geometry.** The solid black boundaries represent the condition $\Phi_M = 0$, i.e. the field is confined in the geometry. The dashed boundaries are the flux conditions for guiding the magnetic field from one quadrant to the next. The gray lines are isopotentials of $\Phi_M$. The dotted rectangle delimits the region of interest (ROI) used in simulations.

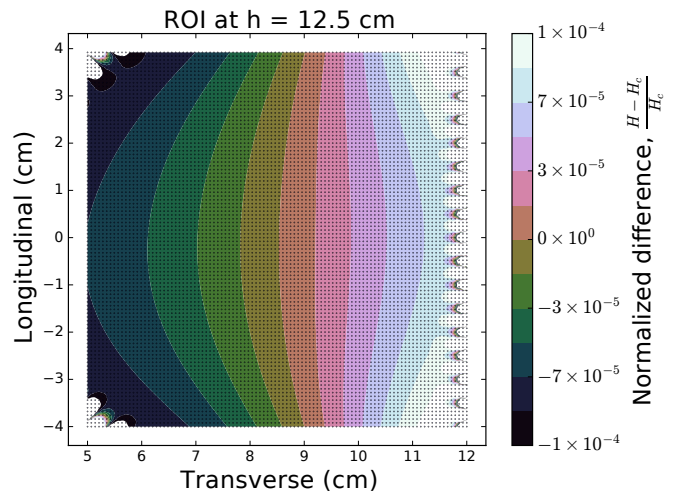

Fig. 3. **Region of interest (ROI).** The dots on the plot show the locations where the magnetic field was calculated using (3). The contour plot is obtained via interpolation of these points. The vertical axis is longitudinal to the main magnetic field component. The horizontal axis has its origin in the center of the coil. The edge effects are due to the wires constituting the coil.

## II. COIL DESIGN

### A. Method

The coil design, shown in Fig.1, confines the field by guiding it through all four quadrants. Each quadrant contains a scalar magnetometer, which measures the averaged field modulus over the sensor volume. A uniform field is required since field gradients broaden the magnetometer's magnetic resonance lines, thus, causing a degradation in sensitivity [16]. The coil was designed using the magnetic scalar potential $\Phi_M$ [17], [18]. In a currentless region, the magnetic field can be expressed as the gradient of a scalar potential $\Phi_M$

$$\vec{H} = -\nabla \Phi_M. \quad (1)$$

In this method, we first define a region representing the coil and then apply flux boundary conditions as demonstrated in Fig.2. The behavior of $\Phi_M$ can be determined by solving the Laplace equation

$$\nabla^2 \Phi_M = 0 \quad (2)$$

for the appropriate set of boundary conditions. The solution is shown on the same figure in terms of isopotentials of $\Phi_M$. Since $\vec{H}$ is always perpendicular to the isopotentials of $\Phi_M$, it can already be seen that the field will be quite uniform in each quadrant.

Next, we simulate the magnetic field produced by this coil. For this, each isopotential is interpreted as a single rectangular current loop. For the present case, this results in a cubic coil of which Fig.2 is the top view. To simulate the behavior of the magnetic field, the Biot-Savart law formulated for a straight wire segment was used [19]

$$\vec{H} = \frac{I}{4\pi} \vec{R}_i \times \vec{R}_f \frac{R_i + R_f}{R_i R_f (R_i R_f + \vec{R}_i \cdot \vec{R}_f)}, \quad (3)$$

where $\vec{R}_{i(f)} = \vec{r} - \vec{r}_{i(f)}$ is the vector from the segment initial (final) point $\vec{r}_{i(f)}$ to the observation point $\vec{r}$. A current loop can then be simulated by composing it of four such wire segments. A Python program was written to simulate the field produced by the entire coil geometry.

### B. Simulations

The following simulation results describe a cubic coil with dimensions $a_o = 250$ mm and $a_i = 1$ mm, as defined in Fig.2. The size of the coil was chosen to fit a 70 mm diameter magnetometer in each quadrant, as shown in Fig.1. Each quadrant consists of 50 current loops and the applied current is 100 mA.

The simulations used a planar region of interest (ROI) in one of the quadrants of the coil, see Fig.2, where field values were evaluated on a grid of 100 by 100 points. Only values of the main field component, hereafter $H$, were used for uniformity characterizations. The transverse components are several orders of magnitude smaller than the longitudinal component. The values for $H$ were then normalized to the field value at the center of the ROI, $H_c$

$$\eta = \frac{H - H_c}{H_c}. \quad (4)$$

The result is shown as a contour plot in Fig.3. Most of the points are distributed narrowly around $\eta = 0$. In order to characterize the uniformity of the field in the ROI, the interquartile range (IQR) of the $10^4$ values for $\eta$ was calculated. The IQR is more robust to outliers, which allows it to cope with the edge effects in Fig.3. The value $\eta_{IQR}$ will be used to characterize the field uniformity in the ROI.

The most important technique for high field uniformity is to properly connect individual current loops. The most uniform field is reached with disconnected individual loops. A simple but realistic case would be to connect the loops at the outer boundary of the coil, the solid black line in Fig.2. This leads to a residual current loop, which follows the mentioned boundary. This can be avoided by connecting the current loops along the dashed boundary in Fig.2. There, the current of one quadrant flows outwards, whereas the current of the neighboring quadrant flows inward. This leads to a compensation of both perturbations and the net effect is reduced.

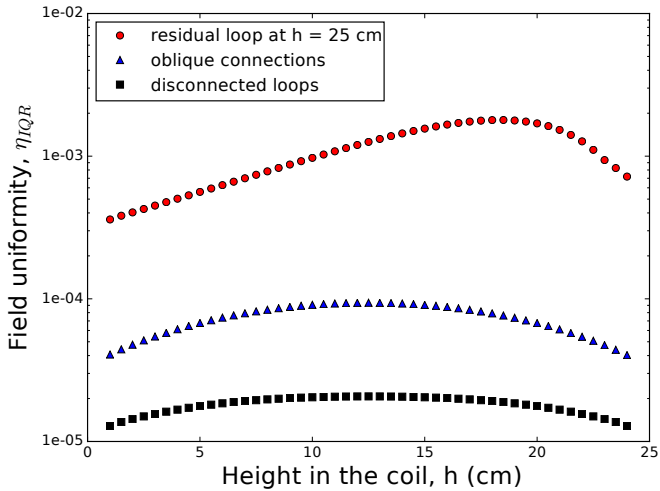

Fig. 4. **Simulation results.** The bottom of the coil lies at 0 cm and the top at 25 cm. Points closer than 1cm to the bottom-, BP, and top-planes, TP, are not included because there the uniformity degrades rapidly. The better uniformity close to BP and TP is a feature of this coil.

We extended the simulations by calculating $\eta_{IQR}$ as a function of the height h of the ROI in the coil. The values of $H_c$, used to calculate $\eta$, have a very uniform dependence on the height h. Thus, Fig.4 shows the field uniformity in each quadrant. One clearly sees that a residual current loop ruins the performance of the coil. Connecting current loops along the inner boundaries of the quadrants leads to no residual current loop, resulting in an order of magnitude better uniformity.

## III. PROTOTYPE

In order to verify the uniformity of this coil design, a prototype was built, see Fig.5. Three different types of PCB panels were used to build it. A total of eight triangular "top panels" form the top and bottom planes of the coil. Four "center panels" were placed between the quadrants. These panels implement the connections between the current loops, see Fig.5b. Finally four "front panels" were connected to the rest of the coil via non-magnetic PCB headers (Molex KK4455 Series). The equidistant wires are 3 mm wide and 35 $\mu$m thick, yielding a total resistivity of 50 $\Omega$ for the entire coil. Each quadrant has 50 current loops, which yields a coil constant $c_B = 249.4 \frac{\text{nT}}{\text{mA}}$. We shall from here on refer to the magnetic flux density $\vec{B}$, which is the quantity that our magnetometers measure.

This coil was tested in a 4-layer mu-metal shield, which saturates around 0.8 T [20]. We measured the uniformity of $B$ using two Cesium magnetometers (CsM) based on free spin precession signals (CsM-FSP) [14]. A CsM measures the average magnitude of the magnetic flux density, $B = ||\vec{B}||$, over the active sensor volume, a 30 mm sphere held at room temperature. This measurement is robust to misalignment of the sensor along the main component of $\vec{B}$. In contrast, a fluxgate sensor measures the flux density along the axis of the sensor. Thus, a slight misalignment of this axis has a noticeable effect on the readout of the measured component.

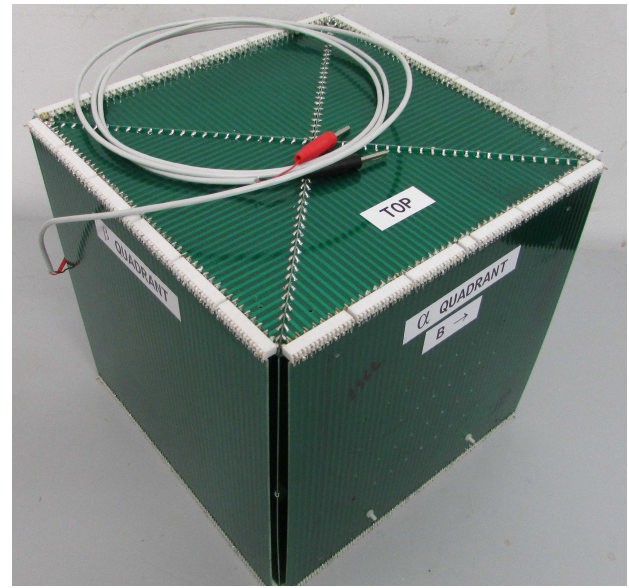

(a)

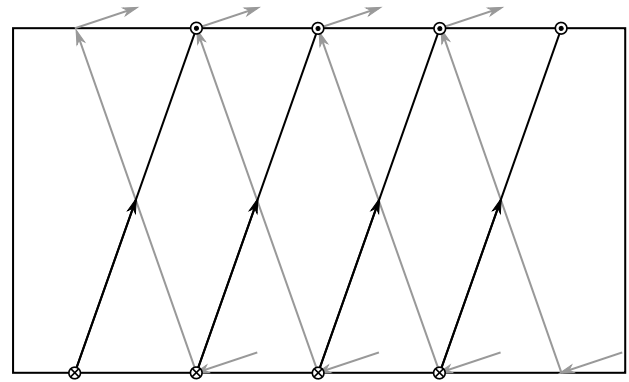

(b)

Fig. 5. **Prototype and wire paths.** a) Prototype with PCBs soldered together to form a cube. The front panels labeled $\alpha$ and $\beta$ quadrant are connected to the rest of the coil via PCB headers and can be removed. The direction of the magnetic flux density $B$ is shown on the $\alpha$ quadrant when applying a positive current. b) Schematic of the current flow on the "center panels". They are double sided PCBs, which implement the connections between current loops, the "oblique connections" case in Fig. 4. The gray lines represent the wires on the backside of the panel, while the black wires are on the frontside.

The uniformity measurements were all done in the $\alpha$ quadrant. Fig.6 shows a schematic of the positions of the two CsM, which were held 5 cm, and 10 cm, apart from each other. The gradiometer returned readings of both sensors simultaneously every 100 ms. Our CsM-FSP gradiometer had a restricted range of operation from 1 $\mu$T to 10 $\mu$T, far from the saturation of the mu-metal.

For each gradiometer configuration a total of 8 measurements were made as a function of the applied magnetic flux density. Between each of these measurements, the mu-metal shield and coil were opened and closed, to get an estimate of the reproducibility of the measurements. The data are represented as the difference in readings, $B_1 - B_2$, as a function of the magnetic flux density in Fig.7. The slope of the linear regression of this data, $s_x$, gives an estimate of the uniformity over the measurement volume, given as the mean $\bar{s}_x$ in Fig.7.

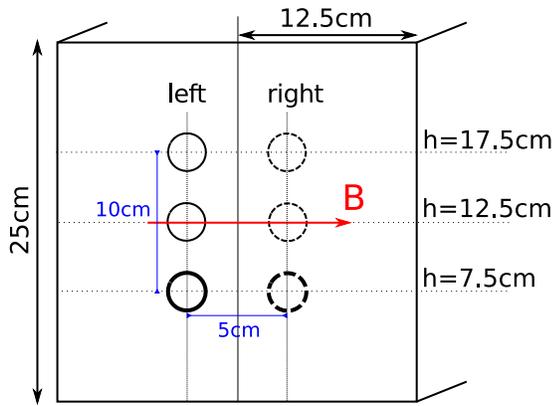

Fig. 6. **Gradiometer locations.** For a horizontal gradiometer, the dashed circles represent sensor 1, reading $B_1$, and the solid circles represent sensor 2, reading $B_2$. For a vertical gradiometer, the thick circles at the bottom represent sensor 1 and the thin circles on top represent sensor 2. Thus measurements were made for 3 horizontal and 2 vertical configurations.

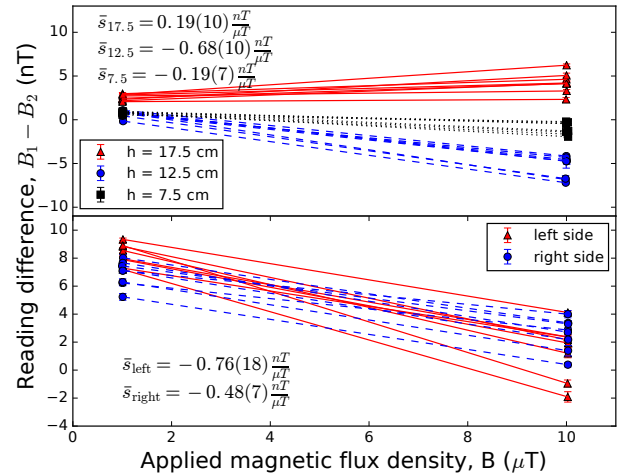

Fig. 7. **Gradiometer results.** For each of the 5 gradiometer configurations 8 measurements were made as a function of the magnetic flux density. The lines are linear regressions of the data points. For each configuration a value $\bar{s}_x$ of the mean of all slopes is given. The value in parenthesis is the standard deviation of those values.

The reproducibility of the measurements was estimated via the standard deviation of the 8 values for $s_x$, given in parenthesis in Fig.7. All $s_x$ are at the sub $10^{-3}$ level of uniformity over the central volume of the $\alpha$ quadrant.

Alternatively, these measurements may be represented in terms of the coil constant, in nT/mA, for each individual sensor. In order to better compare the values, we set the current such that the average reading of all six sensors is 1000 nT, i.e. for I = 4.010 mA. Each sensor records a magnetic flux density value, which is slightly different from the 1000 nT, see Fig.8. The variations are on the sub-nT level for a flux density of 1000 nT. This can again be interpreted as a sub $10^{-3}$ uniformity over the central volume of the $\alpha$ quadrant. These results, as well as the understanding of the working principle of the coil, make it reasonable to assume a similar uniformity in all four quadrants.

The simulation case, which is closest to the measurements, is the "oblique connections" case shown in Fig.4 with the wire paths shown in Fig.5b. However, building the prototype has left some small misalignments between the PCBs, which lead to an imperfect wire pattern compared to Fig.2. This could explain the worse uniformity and the flux density behavior measured in the prototype. The simulations represent the theoretical limit of uniformity which can be reached with this coil design. The results in Figs.7 and 8 show a sub $10^{-3}$ uniformity, which is less than a factor 10 away from the "oblique connections" case in Fig.4. This makes its performance comparable to a $\cos\theta$ coil [4].

The fringes outside of our prototype were measured with fluxgate sensors. A total of four fluxgate sensors were mounted at the outer side of the $\alpha$ quadrant. A "leaking field", of 4 nT to 200 nT, was measured for flux densities, of 1 $\mu$T to 50 $\mu$T, measured inside of the coil, which is typically more than two orders of magnitude smaller. This "leaking field" is due to the gap left by the "front panels", see Fig.5. A design which reduces the size of this gap would improve the confinement aspect. However, small local residual "leaking fields" will always be present.

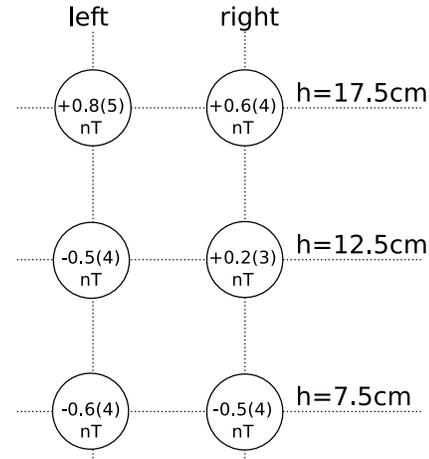

Fig. 8. **Coil constant uniformity.** Set the average reading of all six sensors to be 1000 nT. The coil constant of each individual sensor differs by the amount shown in each circle, representing a CsM. The values in parenthesis are the standard deviation of the coil constant for that sensor.

## IV. CONCLUSION

The presented field confining coil is a simple PCB structure, which uses PCB headers as connectors to open and close the coil. This makes it a viable solution for many applications which require uniform magnetic fields at sub $10^{-3}$ level.

## ACKNOWLEDGMENT

We acknowledge financial support from the FWO Fund for Scientific Research Flanders. Christopher Crawford acknowledges the DOE contract DE-SC0008107. We thank Allard Schnabel and Vira Bondar for useful comments on the manuscript.

## REFERENCES

[1] R. Cacak and J. Craig, "Magnetic field uniformity around near-helmholtz coil configurations," *Review of Scientific Instruments*, vol. 40, no. 11, pp. 1468–1470, 1969.

[2] D. B. Montgomery and R. J. Weggel, *Solenoid magnet design*. Wiley-Interscience, 1969.

[3] J. L. Kirschvink, "Uniform magnetic fields and double-wrapped coil systems: improved techniques for the design of bioelectromagnetic experiments," *Bioelectromagnetics*, vol. 13, no. 5, pp. 401–411, 1992.

[4] A. P. Galvan, B. Plaster, J. Boissevain, R. Carr, B. Filippone, M. Mendenhall, R. Schmid, R. Alarcon, and S. Balascuta, "High uniformity magnetic coil for search of neutron electric dipole moment," *Nuclear Instruments and Methods in Physics Research Section A: Accelerators, Spectrometers, Detectors and Associated Equipment*, vol. 660, no. 1, pp. 147–153, 2011.

[5] N. Nouri and B. Plaster, "Comparison of magnetic field uniformities for discretized and finite-sized standard, solenoidal, and spherical coils," *Nuclear Instruments and Methods in Physics Research Section A: Accelerators, Spectrometers, Detectors and Associated Equipment*, vol. 723, pp. 30–35, 2013.

[6] F. Bertora, A. Viale, E. Molinari, and P. Fabbricatore, "Opening compensation in a 1.5 t open mri magnet for the functional study of the human motor cortex," *IEEE Transactions on Applied Superconductivity*, vol. 20, no. 3, pp. 1831–1834, 2010.

[7] M. G. Abele, "Generation and confinement of uniform magnetic fields with surface currents," *IEEE transactions on magnetics*, vol. 41, no. 10, pp. 4179–4181, 2005.

[8] C. Baker, G. Ban, K. Bodek, M. Burghoff, Z. Chowdhuri, M. Daum, M. Fertl, B. Franke, P. Geltenbort, K. Green *et al.*, "The search for the neutron electric dipole moment at the paul scherrer institute," *Physics Procedia*, vol. 17, pp. 159–167, 2011.

[9] J. Pendlebury, S. Afach, N. Ayres, C. Baker, G. Ban, G. Bison, K. Bodek, M. Burghoff, P. Geltenbort, K. Green *et al.*, "Revised experimental upper limit on the electric dipole moment of the neutron," *Physical Review D*, vol. 92, no. 9, p. 092003, 2015.

[10] N. F. Ramsey, "A molecular beam resonance method with separated oscillating fields," *Physical Review*, vol. 78, no. 6, p. 695, 1950.

[11] S. Afach, G. Bison, K. Bodek, F. Burri, Z. Chowdhuri, M. Daum, M. Fertl, B. Franke, Z. Grujic, V. Hélaine *et al.*, "Dynamic stabilization of the magnetic field surrounding the neutron electric dipole moment spectrometer at the paul scherrer institute," *Journal of Applied Physics*, vol. 116, no. 8, p. 084510, 2014.

[12] V. Y. Shifrin, C. Kim, and P. Park, "Atomic magnetic resonance based current source," *Review of scientific instruments*, vol. 67, no. 3, pp. 833–836, 1996.

[13] D. Budker and M. Romalis, "Optical magnetometry," *Nature Physics*, vol. 3, no. 4, pp. 227–234, 2007.

[14] Z. D. Grujić, P. A. Koss, G. Bison, and A. Weis, "A sensitive and accurate atomic magnetometer based on free spin precession," *The European Physical Journal D*, vol. 69, no. 5, pp. 1–10, 2015.

[15] H. Sasaki, A. Miyajima, N. Kasai, and H. Nakamura, "High-stability dc-current source using nmr lock technique," *IEEE Transactions on Instrumentation and Measurement*, vol. 1001, no. 4, pp. 642–643, 1986.

[16] S. Pustelny, D. J. Kimball, S. Rochester, V. Yashchuk, and D. Budker, "Influence of magnetic-field inhomogeneity on nonlinear magneto-optical resonances," *Physical Review A*, vol. 74, no. 6, p. 063406, 2006.

[17] C. Crawford, "An intuitive and practical interpretation of the magnetic scalar potential," unpublished.

[18] A. E. Marble, I. V. Mastikhin, B. G. Colpitts, and B. J. Balcom, "Designing static fields for unilateral magnetic resonance by a scalar potential approach," *IEEE transactions on magnetics*, vol. 43, no. 5, pp. 1903–1911, 2007.

[19] J. D. Hanson and S. P. Hirshman, "Compact expressions for the biot-savart fields of a filamentary segment," *Physics of Plasmas*, vol. 9, no. 10, pp. 4410–4412, 2002.

[20] W. Randall, "Nickel-iron alloys of high permeability, with special reference to mumetal," *Journal of the Institution of Electrical Engineers*, vol. 80, no. 486, pp. 647–658, 1937.